\documentclass[sigconf]{acmart}
\usepackage{nopageno}
\usepackage[utf8]{inputenc}
\usepackage{booktabs}
\usepackage{multirow}
\usepackage{ctable}
\usepackage{tabularx}
\usepackage{graphicx}
\usepackage{array}
\usepackage[all]{nowidow}
\usepackage{balance}
\usepackage{xspace}
\usepackage{url}
\copyrightyear{2020}
\acmYear{2020}
\setcopyright{acmlicensed}
\acmConference[SIGIR '20]{Proceedings of the 43rd
International ACM SIGIR Conference on Research and Development in
Information Retrieval}{July 25--30, 2020}{Virtual Event, China}
\acmBooktitle{Proceedings of the 43rd International ACM SIGIR Conference on
Research and Development in Information Retrieval (SIGIR '20), July 25--30,
2020, Virtual Event, China}
\acmPrice{15.00}
\acmDOI{10.1145/3397271.3401300}
\acmISBN{978-1-4503-8016-4/20/07}

\begin{document}
\thispagestyle{empty}

\fancyhead{}

\title{Studying Ranking-Incentivized Web Dynamics}

\ccsdesc[500]{Information systems}
\ccsdesc[500]{Information systems~Information retrieval}
\ccsdesc[100]{Information systems~Search engine architectures and scalability}

\keywords{adversarial IR} 

\author{Ziv Vasilisky}
  \email{zivvasilisky@campus.technion.ac.il}
\affiliation{%
  \institution{Technion}
}

\author{Moshe Tennenholtz}
\email{moshet@ie.technion.ac.il}
\affiliation{%
  \institution{Technion}
}

\author{Oren Kurland}
\email{kurland@technion.ac.il}
\affiliation{%
  \institution{Technion}
}

\begin{abstract}
    The ranking incentives of many authors of Web pages play an
    important role in the Web dynamics. That is, authors who opt to
    have their pages highly ranked for queries of interest often
    respond to rankings for these queries by manipulating their pages;
    the goal is to improve the pages' future rankings. Various
    theoretical aspects of this dynamics have recently been studied
    using game theory. However, empirical analysis of the dynamics is
    highly constrained due to lack of publicly available datasets. We
    present an initial such dataset that is based on TREC's ClueWeb09
    dataset. Specifically, we used the WayBack Machine of the
    Internet Archive to build a document collection that contains past
    snapshots of ClueWeb documents which are highly ranked by some
    initial search performed for ClueWeb queries. Temporal analysis of
    document changes in this dataset reveals that findings recently
    presented for small-scale controlled ranking competitions between
    documents' authors also hold for Web data. Specifically,
    documents' authors tend to mimic the content of documents that
    were highly ranked in the past, and this practice can result in improved ranking.
\end{abstract}

\maketitle
\thispagestyle{empty}
\pagestyle{empty}
\renewcommand{\comment}[1]{{\bf {\small !!--#1--!!}}}
\newcommand{\omt}[1]{}
\newcommand{\firstmention}[1]{{\bf #1}}
\newcommand{\myparagraph}[1]{\vspace{0.3\baselineskip}\noindent{\textbf{#1}}.~}

\newcommand{\topical}{Sub-Topic\xspace}
\newcommand{\nonRelevant}{Not-Relevant\xspace}
\newcommand{\docLength}{Doc-Length\xspace}
\newcommand{\incQueryTerms}{Query-Terms\xspace}
\newcommand{\herding}{Herding\xspace}
\newcommand{\biasing}{Biasing\xspace}
\newcommand{\lmart}{LambdaMART\xspace}
\newcommand{\RM}{RM1\xspace}
\newcommand{\control}{Control\xspace}
\newcommand{\group}[1]{Group{#1}\xspace}
\newcommand{\stb}{STB\xspace}
\newcommand{\sth}{STH\xspace}
\newcommand{\nrh}{NRH\xspace}
\newcommand{\dlh}{DLH\xspace}
\newcommand{\qth}{QTH\xspace}

\newcommand{\query}{q}
\newcommand{\topic}{T}
\newcommand{\subTopicParm}[1]{\topic^{sub}_{#1}}
\newcommand{\subTopic}{\topic^{sub}}
\newcommand{\doc}{d}
\newcommand{\corpus}{\mathcal{D}}
\newcommand{\initTag}{{\rm init}}
\newcommand{\topRetGroup}{{\mathcal{D}}_{\initTag}}
\newcommand{\topRetGroupParam}[1]{{\mathcal{D}}^{[#1]}_{\initTag}}
\newcommand{\numDoc}{n}
\newcommand{\arbTerm}{w}
\newcommand{\stopList}{L_s}
\newcommand{\numRelDocs}{l}
\newcommand{\minVal}{\alpha}
\newcommand{\clipVal}{\eta}

\newcommand{\model}{\theta}
\newcommand{\simpleConst}{c}
\newcommand{\queryMLE}{\model^{MLE}_{\query}}
\newcommand{\queryModel}{\model_{\query}}
\newcommand{\anchorParm}{\gamma}
\newcommand{\queryModelSingleParam}[1]{\model_{\query;\anchorParm,\numTerms}^{[#1]}}
\newcommand{\queryModelClipped}[1]{\model_{\query;\numTerms}^{clipped[#1]}}
\newcommand{\queryModelParam}[2]{\theta_{\query}^{[#1;#2]}}
\newcommand{\postQueryModel}{\hat{\theta}_{\query}}
\newcommand{\postDocModel}{\hat{\theta}_{\doc}}
\newcommand{\docModel}{\theta_{\doc}}
\newcommand{\userModel}{\mathcal{U}}
\newcommand{\authModel}{\mathcal{A}}
\newcommand{\utilSymbol}{U}
\newcommand{\util}[2]{\utilSymbol(#1|#2)}
\newcommand{\simFn}{Score}
\newcommand{\reward}{\simFn}
\newcommand{\rewardFunc}[2]{\reward(#1,#2)}
\newcommand{\fbModel}{\mathcal{M}}
\newcommand{\arbSet}{\mathcal{S}_{\subTopicParm{i}}}
\newcommand{\precSetSymbol}{\mathcal{P}}
\newcommand{\precSingle}{\rho}
\newcommand{\precSet}[2]{\precSetSymbol(#1;#2)}
\newcommand{\powSet}{2^{\relDocSet}\setminus \emptyset}
\newcommand{\weight}{w}
\newcommand{\fFunc}{f}
\newcommand{\hFunc}{h}
\newcommand{\relDocSet}{\mathcal{D}_{r}}
\newcommand{\relDocSubset}{D_{r}^{'}}
\newcommand{\nonRelDocSet}{S_{nr}}
\newcommand{\feedbackSetFull}{{\mathcal{F}}^{[\numFB]}}
\newcommand{\feedbackSet}{\mathcal{F}}
\newcommand{\numFB}{k}
\newcommand{\setSize}[1]{|{#1}|}

\newcommand{\ent}{ENT\xspace}
\newcommand{\swOne}{SW1\xspace}
\newcommand{\swTwo}{SW2\xspace}
\newcommand{\coh}{Coherence\xspace}
\newcommand{\minPrec}{MinPrec\xspace}
\newcommand{\precContrib}{PrecContrib\xspace}
\newcommand{\infOne}{InfAP1\xspace}
\newcommand{\infTwo}{InfAP2\xspace}
\newcommand{\infThree}{InfAP3\xspace}
\newcommand{\infFour}{InfAP4\xspace}
\newcommand{\rankParam}[1]{\pi({#1};\topRetGroup)}
\newcommand{\genRanking}{\pi}
\newcommand{\infAP}[1]{\infAPsym(#1)}
\newcommand{\infAPsym}{infAP\xspace}

\newcommand{\set}[1]{\{#1\}}
\newcommand{\definedas}{\stackrel{def}{=}}
\newcommand{\rankEquiv}{\stackrel{rank}{=}}
\newcommand{\kron}[1]{\delta\hspace{-0.2111em}\left[#1\right]}
\newcommand{\kld}[2]{D\left(#1 \; \Big\vert\Big\vert \,\, #2\right)}
\newcommand{\kldSmall}[2]{D\left(#1 \; \vert\vert \,\, #2\right)}
\newcommand{\freq}[2]{{\rm tf}(#1 \in #2)}
\newcommand{\wordIndex}{j} 
\newcommand{\entropy}{H}
\newcommand{\crossEnt}{CE}
\newcommand{\assertion}{\psi} 
\newcommand{\ce}[2]{\crossEnt(#1 \; \vert\vert \,\, #2)}

\newcommand{\prob}{p}
\newcommand{\probhat}{\hat{\prob}}
\newcommand{\modelprob}{\hat{\prob}}
\newcommand{\lmprob}{\modelprob^{LM}}
\newcommand{\ilmprob}{\prob} 
\newcommand{\ilmprobtil}{\tilde{\prob}}
\newcommand{\baseprob}{\prob}
\newcommand{\condArbP}[3]{\ensuremath{#1(#2 \vert #3)}}
\newcommand{\condP}[2]{\condArbP{\prob}{#1}{#2}}
\newcommand{\estCondP}[2]{\hat{\prob}(#1\vert#2)}
\newcommand{\condPhat}[2]{\hat{\prob}(#1|#2)}
\newcommand{\modelcondP}[2]{\condArbP{\modelprob}{#1}{#2}}
\newcommand{\basecondP}[2]{\condArbP{\baseprob}{#1}{#2}}
\newcommand{\lmcondP}[2]{\condArbP{\lmprob}{#1}{#2}}
\newcommand{\genprob}[2]{\inducedprob{\ilmprob}{#1}{#2}} 
\newcommand{\MLE}{MLE\xspace}
\newcommand{\GEN}{GEN\xspace}

\newcommand{\inducedprob}[3]{\ensuremath{#1_{#2}(#3)}}
\newcommand{\inducedlmprob}[2]{\inducedprob{\ilmprob}{#1}{#2}}
\newcommand{\mlprob}[2]{\inducedprob{\ilmprob^{MLE}}{#1}{#2}}
\newcommand{\mlprobTerm}[2]{\inducedprob{\ilmprob^{\,MLE}}{#1}{#2}}
\newcommand{\docinducedlmprob}[1]{\inducedlmprob{\doc}{#1}}
\newcommand{\docinducedlmprobmulti}[1]{\inducedprob{\ilmprob^{Dir}}{\doc}{#1}}
\newcommand{\dirichletParam}{\mu}
\newcommand{\dirichletParamEvalSpecific}{\mu(\evaluationMeasure)}
\newcommand{\dirichletLM}[3]{\inducedprob{\ilmprob^{Dir[#3]}}{#1}{#2}}
\newcommand{\numTerms}{\delta}
\newcommand{\mixLM}[2]{\inducedprob{\ilmprob^{Mix[\mixParam,\numTerms]}}{#1}{#2}}
\newcommand{\mixLMParam}[4]{\inducedprob{\ilmprob^{Mix[#3,#4]}}{#1}{#2}}
\newcommand{\dirichletLMTerm}[3]{\dirichletLM{#1}{#2}{#3}}
\newcommand{\bareMLProb}[1]{\ilmprob^{MLE}_{#1}}
\newcommand{\bareMLProbTerm}[1]{\widetilde{\ilmprob}^{\,MLE}_{#1}}
\newcommand{\bareDirProb}[1]{\ilmprob^{[\dirichletParam]}_{#1}}
\newcommand{\docLengthMax}{\psi}

\newcommand{\ourAlgSym}{Subsets}
\newcommand{\ourAlg}[1]{\ourAlgSym(#1)}
\newcommand{\ourUAlg}[1]{U\ourAlgSym(#1)}
\newcommand{\ourUUAlg}[1]{U\ourAlgSym(#1)}
\newcommand{\select}[1]{Select(#1)}
\newcommand{\svmRank}{SVM$^{rank}$\xspace}
\newcommand{\relModel}{R\xspace}
\newcommand{\mixModel}{MM\xspace}
\newcommand{\swlm}{SWLM\xspace}
\newcommand{\medmm}{MEDMM\xspace}
\newcommand{\prm}{Passage\xspace}
\newcommand{\predm}{PredictM\xspace}
\newcommand{\svmc}{SVMC\xspace}
\newcommand{\cltr}{ClustLTR\xspace}
\newcommand{\cmrf}{ClustMRF\xspace}
\newcommand{\drift}{ClustDrift\xspace}
\newcommand{\init}{Initial\xspace}

\newcommand{\ap}{AP\xspace}
\newcommand{\robust}{ROBUST\xspace}
\newcommand{\govtwo}{GOV2\xspace}
\newcommand{\clueweb}{ClueWeb\xspace}

\newcommand{\map}{MAP\xspace}
\newcommand{\precFive}{p@5\xspace}
\newcommand{\ndcg}{NDCG\xspace}
\newcommand{\ablation}{Exclude $\fFunc_i$\xspace}
\newcommand{\single}{Only $\fFunc_i$\xspace}
\newcommand{\rimap}{RI(MAP)\xspace}
\newcommand{\rindcg}{RI(NDCG)\xspace}

\newcommand{\statSymbolAll}{\star}
\newcommand{\statSymbolRM}{r}
\newcommand{\statSymbolSubset}{s}
\newcommand{\statSymbolFB}{m}
\newcommand{\negWeight}{\;\;\,}

\newcommand{\minP}{Min\xspace}
\newcommand{\ameanPrec}{AMean\xspace}
\newcommand{\gmeanPrec}{GMean\xspace}

\newcommand{\hypA}{HypoA\xspace}
\newcommand{\hypB}{HypoB\xspace}
\newcommand{\hypC}{HypoC\xspace}

\newcommand{\figHeight}{5cm}
\newcommand{\figWidth}{8.1cm}

\newcommand{\queryCover}{QueryCover\xspace}
\newcommand{\fracQuery}{FracQuery\xspace}
\newcommand{\pass}{P\xspace}
\newcommand{\act}{A\xspace}
\newcommand{\stba}{\stb-\act}
\newcommand{\stha}{\sth-\act}
\newcommand{\stbp}{\stb-\pass}
\newcommand{\sthp}{\sth-\pass}

\section{Introduction}
The Web is a dynamic retrieval setting. An important part of this
dynamics is due to the ranking incentives of Web pages' authors. That
is, some are interested in having their pages highly ranked for
queries of interest. As a result, they might respond to induced
rankings by modifying their documents so as to improve their future
ranking --- a practice also known as search engine optimization
\cite{Gyongyi+Molina:05a}.

Recently, there has been a growing research interest in analyzing the
dynamics just described using game theory
\cite{Tennenholtz+Kurland:19a}\footnote{Similar dynamics in recommendation systems was studied in \cite{BenPorat+Tennenholtz:18a}.}. For example, it was shown that the
probability ranking principle \cite{Robertson:77a} is not optimal in
competitive retrieval settings as it leads to a reduced topical
diversity in the corpus \cite{Basat+al:17a}. In addition, the document
manipulation strategies of documents' authors were analyzed using game
theory \cite{Raifer+al:17a}. The robustness of induced rankings under adversarial document modifications has also been explored
\cite{Goren18}.

However, it seems that to empirically analyze the ranking incentivized
dynamics of the Web, access to the query logs of commercial search
engines is called for \cite{Tennenholtz+Kurland:19a}. Even with such
access allowed, isolating and studying specific aspects of the
dynamics can be a difficult challenge
\cite{Tennenholtz+Kurland:19a}. As a result of this state-of-affairs,
there have been recent small-scaled controlled studies of ranking
competitions between incentivized documents' authors
\cite{Raifer+al:17a}. Yet, increasing the scale of such studies and performing analysis of real Web dynamics driven by induced rankings still remains a challenge \cite{Tennenholtz+Kurland:19a}.

To address the challenge(s) just mentioned, we have developed a novel
dataset available at
\url{https://github.com/hscw09dataset/hscw09}. Specifically, we used
the WayBack Machine of the Internet Archive
(\url{https://archive.org/web/}) to find past versions of a selected
set of documents in TREC's ClueWeb09 collection; this is a Web crawl
from 2009 which serves as a standard testbed for Web retrieval
evaluation. The selected documents were the most highly ranked in
response to ClueWeb09 queries.

Analysis of the dataset reveals that some characteristics of the
temporal changes of ClueWeb09 documents are in accordance with those
which have emerged from analysis of small-scaled controlled ranking
competitions between students who authored short plaintext documents
\cite{Raifer+al:17a}. Thus, while we have no concrete evidence that
the authors of these documents were incentivized to rank-promote them
for ClueWeb09 queries, aspects of the actual modifications do conform
to those applied in a competitive ranking setting by rank-incentivized
documents' authors. A case in point, documents which were promoted to
the highest rank for a given query became more similar to those highly
ranked for the query in the past. The rationale for this competitive
modification strategy in a setting with undisclosed ranking function
whose induced rankings can be observed was recently provided using a
game theoretic analysis \cite{Raifer+al:17a}. This theoretical
finding was supported by the analysis of 
controlled ranking competitions between students \cite{Raifer+al:17a}.

We believe that the dataset we have developed, and its future
potential developments, will help to better understand and further explore Web dynamics
with respect to the inherent competitive retrieval setting which is
driven by rank-incentivized authors.

\section{Related Work}
Most past work on adversarial IR has focused on spamming
\cite{Castillo+Davison:10a}. Search engine optimization needless
necessary be black-hat or spamming \cite{Gyongyi+Molina:05a}. In fact, legitimate (a.k.a. white hat) document modifications and rank-promotion strategies play an important role in driving the dynamics of the Web retrieval setting \cite{Gyongyi+Molina:05a}. One of our goals in developing the dataset is to study this white hat dynamics.

There is work on predicting and analyzing changes of Web
pages; e.g., \cite{Radinsky+Bennett:13a,Radinski+al:13a}. These studies are for general dynamics and not for ranking-oriented dynamics which we address in this paper.

There has also been work on using the past versions of a Web page so as to improve its current representation for ranking purposes \cite{Ablimit10,Elsas10,Nunes11}. Analysis of temporal document changes as that we present here was not reported. Still, we believe that our dataset will help to facilitate and further this line of research as well.

\section{The Dataset}
\label{sec:dataset}
Our first goal was to create a dataset that allows to study the temporal
dynamics of documents that are highly ranked for queries. To this end,
we used TREC's ClueWeb09 dataset \cite{ClueWeb09}. The titles of
topics $1$-$200$ served as queries. Krovetz stemming was applied to
queries and documents.
We applied standard
language-model-based retrieval for each query. Specifically, we used
the KL divergence between the unsmoothed maximum likelihood estimate
induced from the query and the Dirichlet smoothed document language models for
ranking; the smoothing parameter was set to $1000$
\cite{Lafferty+Zhai:01a}. We scanned each retrieved list top down and
removed documents whose Waterloo's spam score \cite{Waterloo10} was
below $50$ until we have accumulated $50$ documents. We use $L^{\query}_{0}$ to denote the retrieved list for query $\query$; $0$ is the timestamp. The set of all documents in the lists retrieved for the $200$ queries is referred to as the {\em base set}; it contains $9986$ different documents. (Some documents are among the top-retrieved for more than one query.)

The next step was to collect past versions of the documents in
$L^{\query}_{0}$ for each query $\query$. To that end, we used the
WayBack Machine of the Internet Archive
(\url{https://archive.org/web/}). The Internet Archive is a large
repository of Web pages crawlled in different points in time. Since
ClueWeb09 is a crawl from 2009, we used past versions from
2008. Specifically, we used $12$ time intervals, each corresponds to a
calendar month in 2008. We used a document's
{\em earliest} snapshot in an interval as its version for the interval in case a snapshot
existed. Thus, we consider at most $13$ versions of a document: the
ClueWeb version (\# 0) and the 12 past versions from 2008. The 2008
document lists for query $\query$ are denoted $L^{\query}_{-1},
\ldots, L^{\query}_{-12}$ where $-1$ corresponds to the December
interval and $-12$ corresponds to the January interval. The number of documents in each of these lists ($\le 50$) depends on the number of documents from $L^{\query}_{0}$ with at least one past snapshot in the corresponding interval. In what follows, we use $\doc^{\query}_{i}$ where $i \in \{0,-1,\ldots,-12\}$ to denote a document in list $L^{\query}_{i}$.

Out of the $9986$ different documents in the base set, $7425$
($74.4\%$) have at least one past version in one of the twelve 2008
time intervals. The mean number of past versions of a document
is $3.06$ and the median is $2$. For documents with at least one past version, the mean is $4.1$ and the median is $3$.


\section{Temporal Document Changes}
\label{sec:docChange}
\begin{figure}[t]
  \hspace*{-.2in}
  \includegraphics[width=11cm,height=5.5cm]{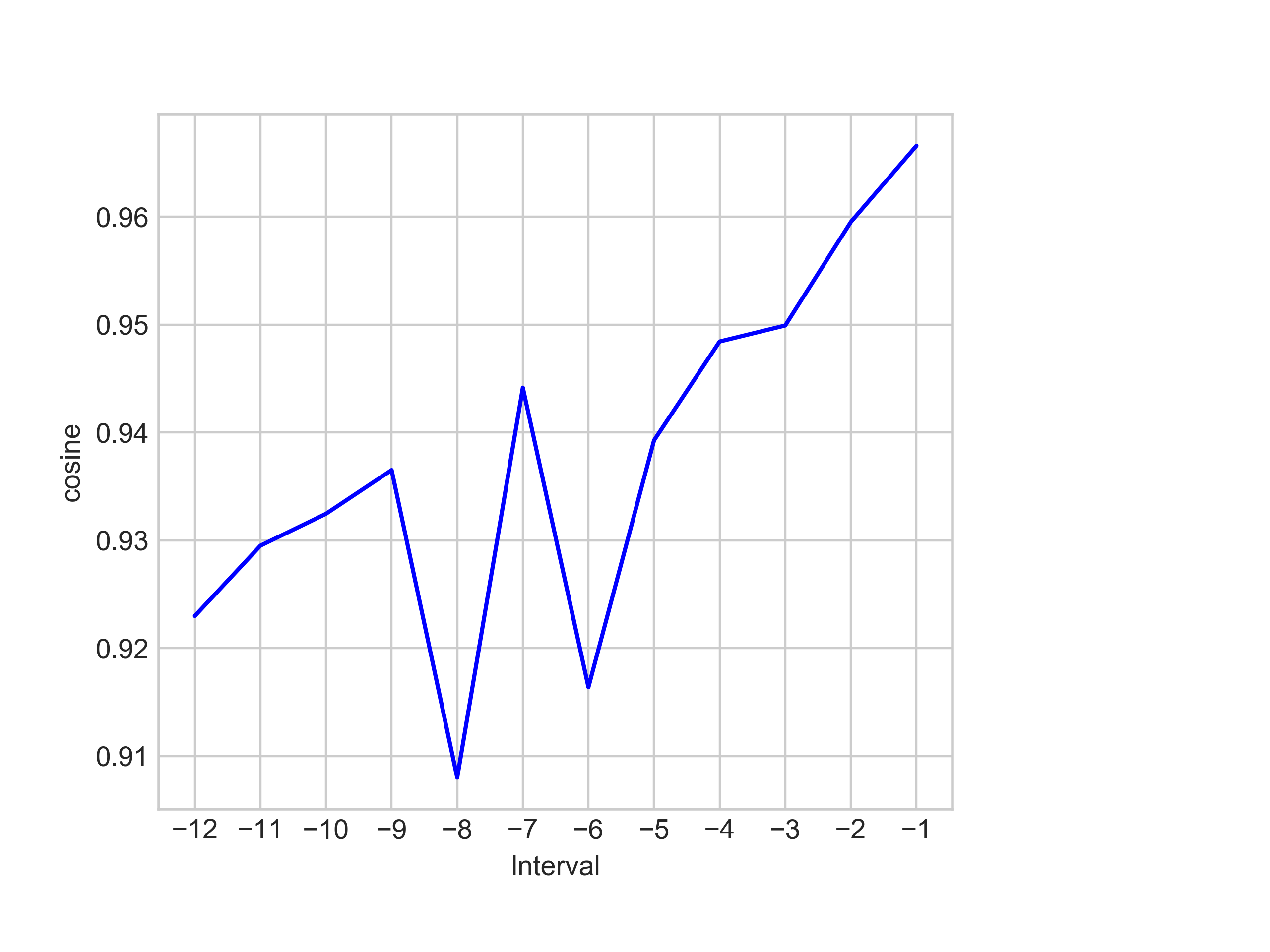}
  \caption{\label{fig:sim} Average cosine similarity between a
    document at time interval $0$ (i.e., the ClueWeb09 document) and its past version (if exists) at time interval $i$.}
\end{figure}

Figure \ref{fig:sim} presents the average similarity between documents in the base set and their past versions. Specifically, for each time interval, we compute the average cosine between the tf.idf vectors representing documents available in the interval and those representing their $0$ versions (i.e., the original ClueWeb09 versions). As could be expected, we see in Figure \ref{fig:sim} a general upward trend: the past versions of documents gradually become more similar (with a few exceptions) to their $0$ versions. 

We next analyze some characteristics of the changes of documents in
our dataset along time. Our analysis is
inspired by the ranking competitions that Raifer et
al. \cite{Raifer+al:17a} organized between students. The students
wrote plaintext documents of up to 150 terms and manipulated them
along time so as to have them highly ranked for queries of interest by
an unknown ranking function. Raifer et al. \cite{Raifer+al:17a} found
that the documents had an increasing number of query term occurrences
along time, a reduced number of stopwords, and a decreased term
distribution entropy which attests to some extent to reduced content breadth \cite{Bendersky2011}.

We use a few definitions.
The ratio of query terms that appear in
a document is referred to as \firstmention{QueryTermsRatio}. The relative {query terms ratio} of document
$\doc^{\query}_{i}$ with respect to its $0$ version, $\doc^{\query}_{0}$, is denoted $\mathbf{rqtr(\doc^{\query}_{i})}$.
The ratio of stopwords to non-stopwords in a document is denoted \firstmention{StopwordsRatio}\footnote{We used the NLTK (\url{https://www.nltk.org/}) stopword list.}. The relative {stopwords ratio} with respect to version $0$ of document
$\doc^{\query}_{i}$ is denoted $\mathbf{rswr(\doc^{\query}_{i})}$.
High stopword occurrence in documents was shown to be an effective document relevance prior which corresponds to the premise that stopword presence attests to document quality \cite{Bendersky2011}. The entropy of a document's term distribution, henceforth Entropy, potentially attests to its content breadth \cite{Bendersky2011}. The relative {entropy} of document $\doc^{\query}_{i}$, with respect to its $0$ version, is denoted $\mathbf{rent(\doc^{\query}_{i})}$.

\begin{figure}[t]
  \includegraphics[width=\linewidth,height=5cm]{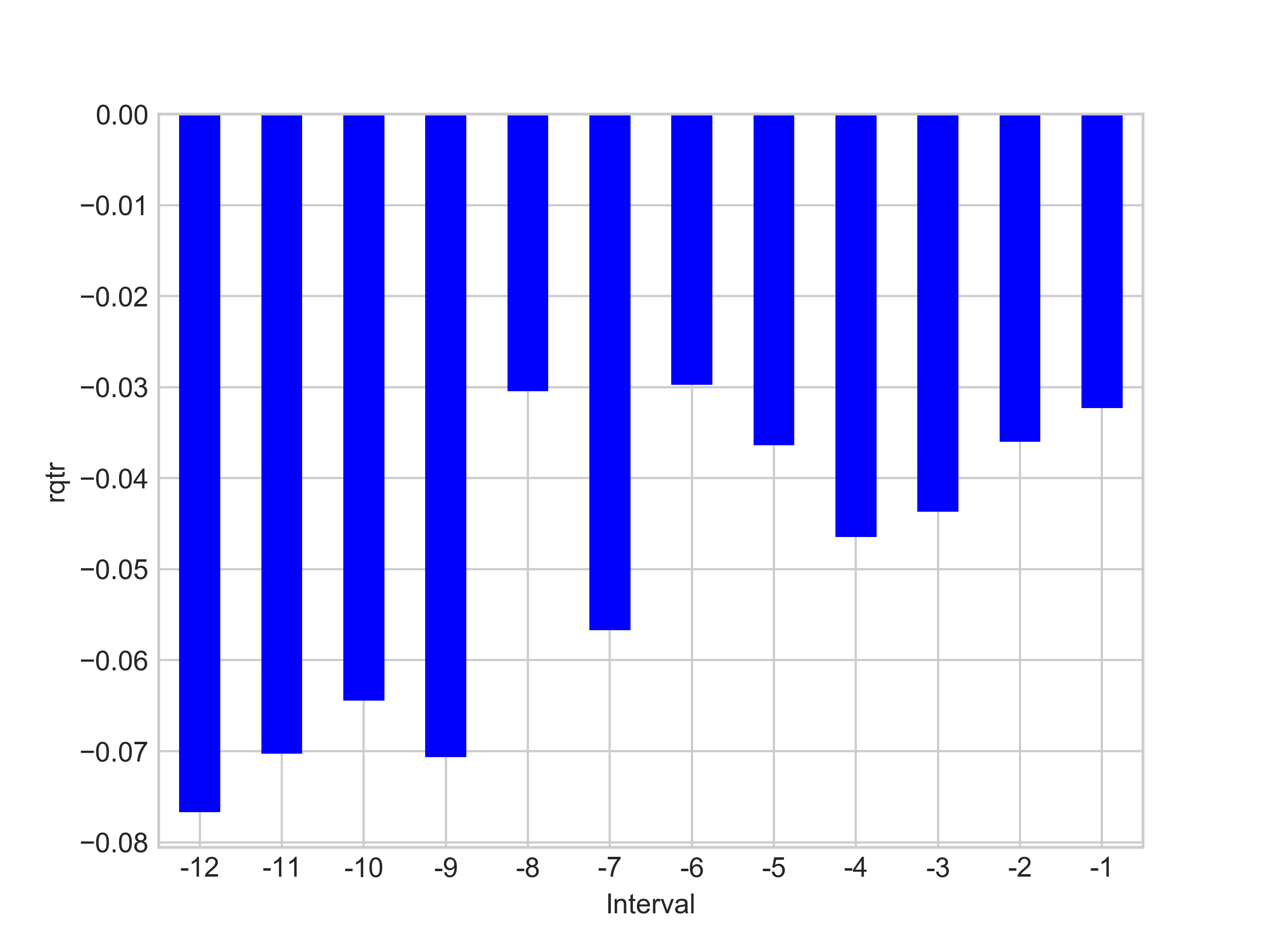}
  \caption{\label{fig:qtr} The average relative query terms ratio ($rqtr$).}
\end{figure}

Figure \ref{fig:qtr} shows that all the average (per interval)
relative query-term ratios for documents are negative; i.e., past versions of the ClueWeb09 documents have lower query terms ratio. Furthermore, we see an overall upward trend which attests to increased use of terms of queries for which the documents were highly ranked. Thus, it seems as if there is some ``effort'' in promoting in rankings these documents for the ClueWeb09 queries although there is clearly no explicit evidence. We note that this trend was also observed in the ranking competitions of Raifer et al. \cite{Raifer+al:17a} where students explicitly worked on rank-promoting their documents for given queries.

\begin{figure}[t]
  \includegraphics[width=\linewidth,height=5cm]{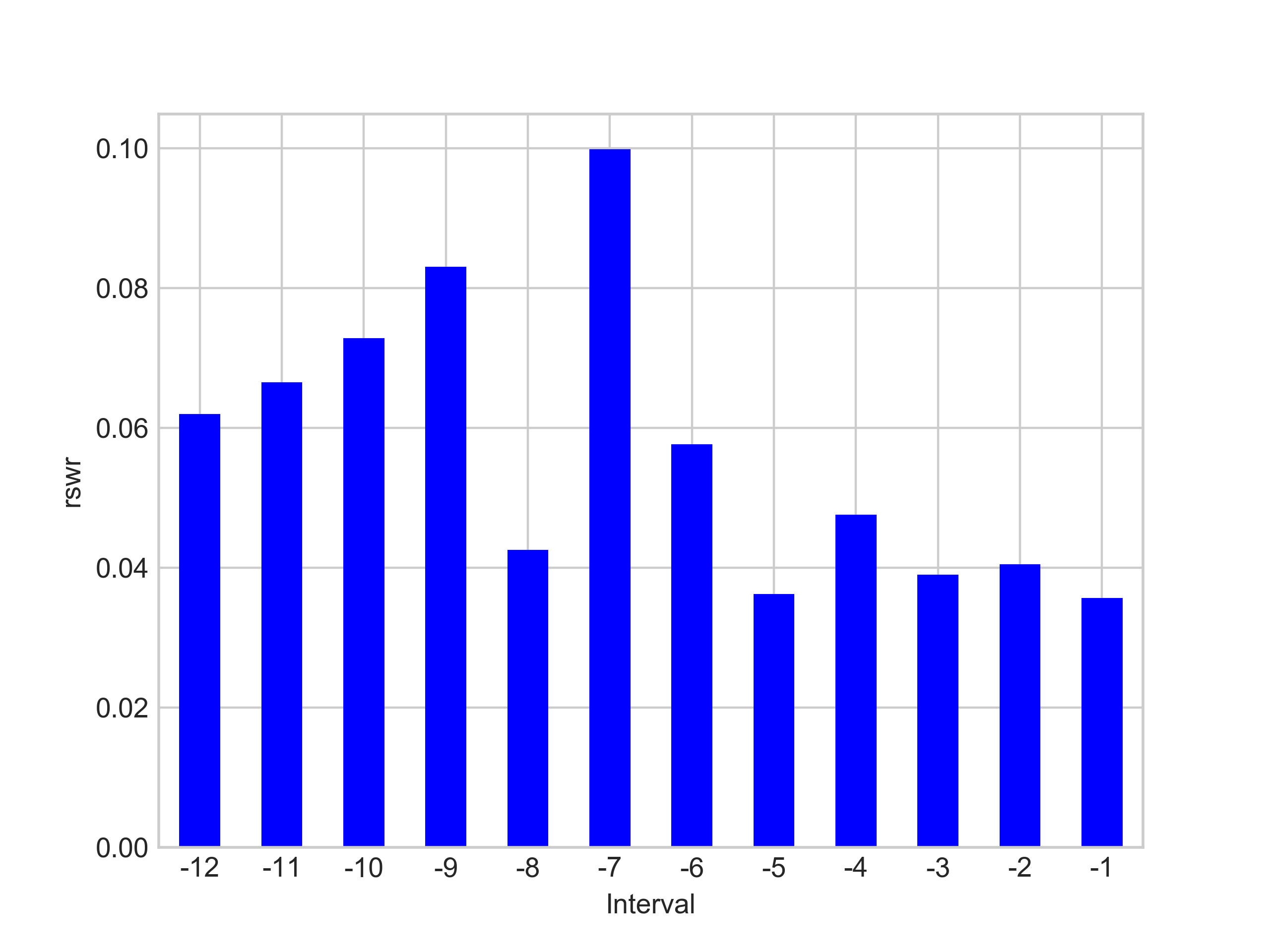}
  \caption{\label{fig:swr} The average relative stopword ratio ($rswr$).}  
\end{figure}

We see in Figure \ref{fig:swr} that the average, per-interval, relative stopword ratio ($rswr$) is positive for all intervals. That is, the stopword ratio of past versions of the ClueWeb09 documents was higher than that for the ClueWeb09 versions. We also observe in Figure \ref{fig:swr} a general downward trend from interval -7 (June 2008) to interval -1 (December 2008). This finding echoes those from Raifer et al.'s \cite{Raifer+al:17a} controlled ranking competitions with plaintext documents.

\begin{figure}[t]
  \includegraphics[width=\linewidth,height=5cm]{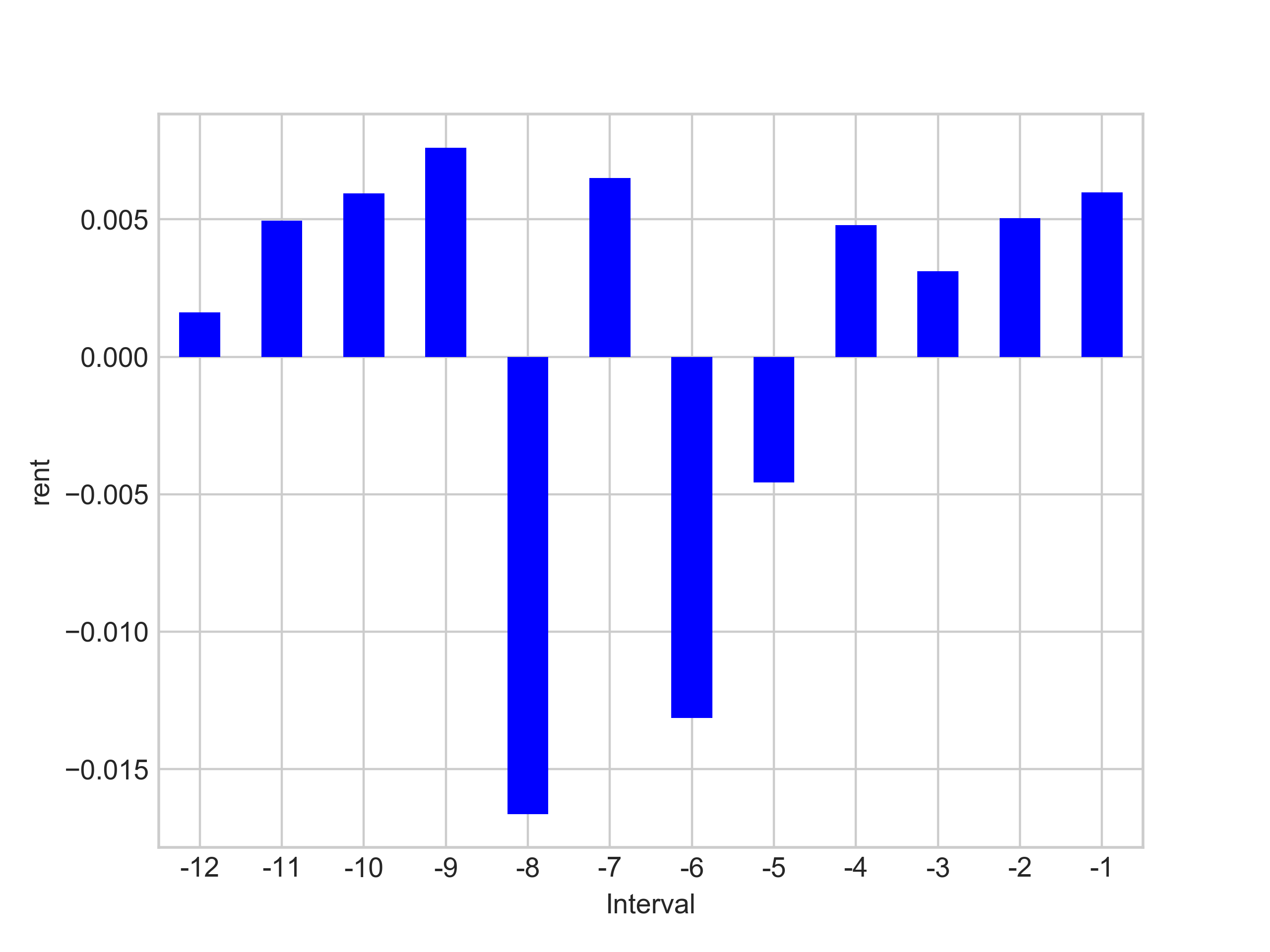}
  \caption{\label{fig:entropy} The average relative entropy ($rent$).}
\end{figure}

Figure \ref{fig:entropy} shows that for most intervals, the past
versions of ClueWeb09 documents had higher entropy than that for the
ClueWeb09 versions. This entropy drop (between the past and the
ClueWeb09 version), which attests to decreased
content breadth, is in line with Raifer et al.'s findings \cite{Raifer+al:17a}.

To summarize, although our dataset is
composed of real Web documents, and we have no explicit evidence
that their authors were actually trying to promote them for ClueWeb09
queries, the temporal changes of the documents are similar in
several respects to those observed for short plaintext documents used
in controlled ranking competitions with explicit rank-promotion
incentives of their authors \cite{Raifer+al:17a}.

\section{Documents' temporal changes from a ranking perspective}
\label{sec:rankChanges}
We next analyze some changes of rankings throughout the time intervals. In addition, we study the
documents' changes from a ranking perspective. Recall that we use standard language-model-based
retrieval. Our goal is to focus on content changes of documents
rather than explore changes to anchor text, hyperlinks, etc.

\begin{figure}[t]
  \hspace*{-.3in}
  \includegraphics[width=10cm,height=6cm]{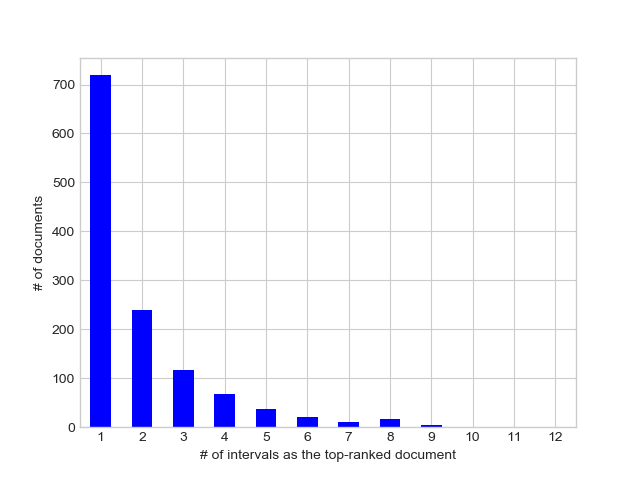}
  \caption{\label{fig:winChanges} The number of documents in the base
    set whose past versions are the highest ranked for $x$ intervals.}
\end{figure}

In Figure \ref{fig:winChanges} we present the number of documents
in the ClueWeb09 base set whose past versions are ranked at the highest rank for
$x$ intervals. We see that very few documents have past versions that are positioned at the
highest rank for more than very few intervals. In other words, there
is significant dynamics in terms of the highest rank along time. This dynamics is reminiscent of that observed by Raifer et al. \cite{Raifer+al:17a} in their controlled ranking competitions.

\myparagraph{Strategic manipulations?}  Raifer et
al. \cite{Raifer+al:17a} analyzed --- theoretically and empirically
--- the document manipulation strategies employed by students in their
controlled competitions. As noted above, we have no explicit evidence
that authors of ClueWeb09 pages were actually incentivized to
rank-promote their documents for the ClueWeb09
queries. Nevertheless, the findings we presented above with regard to the characteristics of documents' changes attested to the
fact that the manipulations of the documents along time did correspond to those observed in Raifer et al.'s competitions for rank-incentivized authors. Hence, we now further address the characterization of the manipulations from a ranking perspective using the analysis proposed by Raifer et al. \cite{Raifer+al:17a}.

We term the document ranked at the highest rank at some time interval
the {\em winner} of the interval; all other documents in the interval
are referred to as {\em losers} \cite{Raifer+al:17a}. We study two
groups of documents: winners (\firstmention{W}) of intervals and those which were losers (\firstmention{L}) for
at least three consecutive intervals before a current interval in
which they became winners. We contrast the two
groups with respect to four features. The first three are those used in Section \ref{sec:docChange} : QueryTermsRatio, StopwordsRatio and Entropy. The
fourth feature is the similarity to the previous winner
(\firstmention{SimPW}): the cosine similarity between the tf.idf vector
representing the document and the vector representing the document
which was the highest ranked in the previous interval. Following
Raifer et al.'s analysis \cite{Raifer+al:17a}, we further divide the
L group to two with respect to each feature: those whose
feature value three intervals before winning was less or equal (L$\leq$W) or higher (L$>$W) than that
of the winner of that interval. 

\begin{figure}[t]
  \hspace*{-.3in}
    \includegraphics[width=10.5cm,height=7.5cm]{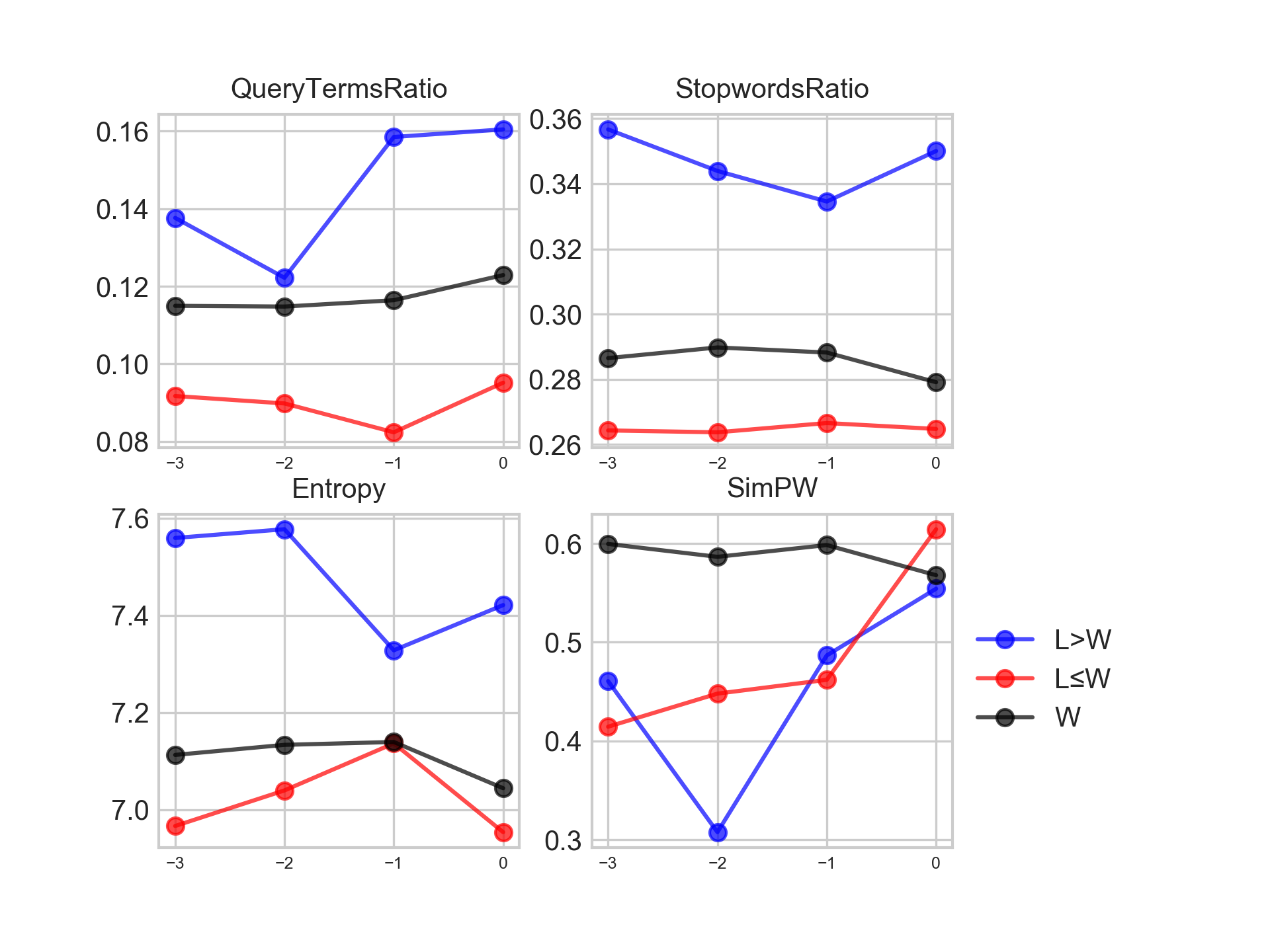}
  \caption{\label{fig:winlose} Average feature values per time
    interval: Contrasting winner documents (W) with documents which
    were losers (L) for at least three consecutive time intervals
    before becoming winners. \textbf{L$>$W} and \textbf{L$\leq$W} mark
    losers whose feature value was higher, or less or equal to, that
    of the winner three time intervals before becoming a winner.}
\end{figure}

We see in Figure \ref{fig:winlose} that the values of the QueryTermsRatio,
StopwordsRatio and Entropy features remain relatively stable for winners. 
In contrast, these feature values for the two groups of losers who became winners can quite fluctuate. More importantly, we observe an upward trend of SimPW for both groups of losers. This means that losers were making their documents more similar to those of winners from the previous interval. This strategy was theoretically motivated for rank-incentivized authors by using a game theoretic modeling of the competitive retrieval setting \cite{Raifer+al:17a}. Furthermore, the empirical finding is the same as that of Raifer et al.'s \cite{Raifer+al:17a} for their controlled ranking competitions.

\section{Conclusions and Future Work}
We described a dataset we have created which contains past versions of
ClueWeb09 documents that are highly ranked for ClueWeb09
queries. Analysis of the dataset revealed that aspects of
temporal changes of documents are in accordance with those reported
for small-scale controlled ranking competitions between students who
authored and manipulated short plaintext documents
\cite{Raifer+al:17a}.

We plan to further develop this dataset and use it to extend our study of the Web dynamics which revolves around rankings induced for queries.

\myparagraph{Acknowledgments}
We thank the reviewers for their comments.
The work by Moshe Tennenholtz was supported by funding from the European Research Council (ERC) under the European Union’s Horizon 2020 research
and innovation programme (grant agreement 740435).

\bibliographystyle{ACM-Reference-Format}
\bibliography{main}
\end{document}